\documentclass[aps,pre,superscriptaddress,twocolumn,amsmath,amssymb,showpacs]{revtex4}
\usepackage{graphicx}
\usepackage{xfrac} 

\usepackage{xcolor}

\begin{document}

\title{Most probable paths in temporal weighted networks: \\An application to ocean transport}

\author{Enrico Ser-Giacomi}
\affiliation{IFISC, Instituto de F\'isica Interdisciplinar y
Sistemas Complejos (CSIC-UIB), E-07122 Palma de Mallorca,
Spain}
\author{Ruggero Vasile}
\affiliation{IFISC, Instituto de F\'isica Interdisciplinar y
Sistemas Complejos (CSIC-UIB), E-07122 Palma de Mallorca,
Spain}
\affiliation{Ambrosys GmbH, Albert-Einstein-Str. 1-5 , 14473 Potsdam, Germany}
\author{Emilio Hern\'{a}ndez-Garc\'{\i}a}
\affiliation{IFISC, Instituto de F\'isica Interdisciplinar y
Sistemas Complejos (CSIC-UIB), E-07122 Palma de Mallorca,
Spain}
\author{Crist\'obal L\'{o}pez}
\affiliation{IFISC, Instituto de F\'isica Interdisciplinar y
Sistemas Complejos (CSIC-UIB), E-07122 Palma de Mallorca,
Spain}

\date{\today}


\begin{abstract}

We consider paths in weighted and directed temporal networks,
introducing tools to compute sets of paths of high probability.
We quantify the relative importance of the most probable path
between two nodes with respect to the whole set of paths, and
to a subset of highly probable paths which incorporate most of
the connection probability. These concepts are used to provide
alternative definitions of betweenness centrality. We apply our
formalism to a transport network describing surface flow in the
Mediterranean sea. Despite the full transport dynamics is
described by a very large number of paths we find that, for
realistic time scales, only a very small subset of high
probability paths (or even a single most probable one) is
enough to characterize global connectivity properties of the
network.
\end{abstract}

\maketitle

\section{Introduction}

Many real-world systems can be studied by using the network
paradigm~\cite{newman2009networks, newman2006structure}. Though
in many of these situations the connections between network
nodes are constant in time and it is sufficient to adopt a
static network description, this approach may not be always
suitable and a temporal network description is
required~\cite{holme2012temporal}. Relevant examples are
epidemic spreading, human communication or transportation
networks, i.e. systems where the strong time variability plays
a crucial role in determining connections and
interactions~\cite{masuda2013predicting,pan2011path,tang2009temporal}. Also, approaching continuous dynamical
systems from a network perspective
\cite{santitissadeekorn2007identifying,gfeller2007complex}
requires a spatio-temporal discretization that, in the
non-autonomous case, often determines a marked time-dependence
of the resulting networks. This feature is clearly
recognizable, in particular, in climate networks
\cite{yamasaki2008climate,deza2013inferring,mheen2013interaction,stolbova2014topology}
and in networks describing connectivity by fluid flows~
\cite{rossi2014hydrodynamic,sergiacomi2015flow}.

Prominent connectivity patterns in networks can be revealed by
introducing the concept of paths~\cite{newman2009networks,
newman2006structure}. While its definition is simple and
intuitive for static or aggregated networks, for the temporal
case paths between nodes can suddenly appear and disappear in
time~\cite{holme2012temporal,kempe2000connectivity,
kostakos2009temporal, tang2010small, tang2010analysing,
kim2012temporal, starnini2012random,lentz2013unfolding,masuda2013temporal}. Thus, there is recently
a focus towards the definition and characterization of paths in
temporal networks. The concept of shortest path in static
network analysis has been generalized to include information on
the time necessary to establish a space-time connection between
nodes. This was the motivation behind the development of
fastest path analysis, specially for unweighted and undirected
networks \cite{holme2012temporal,starnini2012random}. Although
it is relevant to study the time required to build a path among
two nodes, it is equally crucial to understand how to quantify
the importance and the distribution of such paths. This issue
becomes essential when one tries to exploit this information in
order to define and evaluate global network properties, for
example betweenness centrality.

In this Letter we extend the concept of most probable
path (MPP)~\cite{braunstein2003optimal,sevon2006link,gautreau2007arrival,gautreau2008global,brockmann2013hidden}
to the case of temporal, weighted and
directed networks.
We quantify the relative importance of the MPP with respect to
the whole set of paths, and to the subset of highly probable
paths (HPPs) incorporating most of the connection probability.
Using such sets of paths we are able to define a betweenness centrality measure for temporal weighted networks.
The approach presented here
is applied to a flow network describing ocean water transport
by surface currents in the Mediterranean Sea.
In this example
we demonstrate that information contained in MPPs (or in small subsets
of HPPs) suffices to describe all mayor transport properties, despite the
number of such paths being just a very small fraction of the
full set.
The MPPs correspond to the main carriers of ocean mass transport
(showing connectivity patterns)
and MPP-betwenness to a measure for a clear identification of
the main avenues of water transport in the Mediterranean Sea.
%

\section{Most probable paths}

The analysis is restricted to
a time interval $[t_0, t_{M}]$ in which $M+1$ snapshots of the state of the network are
taken at times $t_l\, =t_0+l\tau$, $l=0,1,...,M$, with $\tau$
the time between them. We consider a temporal, directed and
weighted network of $N$ nodes. Its time-dependent connectivity
is described by a set of weighted adjacency matrices
$\mathbf{A}^{(l)}$, $(l = 1,..,M)$, in which the matrix element
$\mathbf{A}^{(l)}_{IJ}\geq 0$ specifies the strength of
connectivity from $I$ to $J$ during the time interval
$[t_{l-1}, t_l]$. A convenient way to analyze the system is
using \emph{time-ordered graphs} (TOGs) \cite{kim2012temporal}.
Formally, the TOG can be considered a static network of
$N\times (M+1)$ nodes with directed and causal links. For each
snapshot $l$, a group $V(t_l)$ of $N$ nodes replicating the
nodes of the original network can be defined. Links are then
established only from nodes at successive times, i.e. from
$i_{l-1} \in V(t_{l-1})$ to $j_{l}\in V(t_l)$ with the weights
given by those in the original temporal network:
$\mathbf{A}^{(l)}_{i_{l-1}~j_{l}}$.


We now consider a flow or transport process by releasing
independent random walkers in each node of the network. Their
motion is assumed to be Markovian and is defined by single-step
transition probabilities proportional to the entries in the
adjacency matrices. Specifically the probability of reaching
node $k_{l}$ at time $t_l$ under the condition of being at
$k_{l-1}$ at time $t_{l-1}$ is:
\begin{equation}
\mathbf{T}^{(l)}_{k_{l-1}
k_{l}} \equiv \frac{\mathbf{A}^{(l)}_{k_{l-1},k_{l}}}{
s_{out}^{(l)}(k_{l-1})} \ .
\end{equation}
Here $s_{out}^{(l)}(k) = \sum_{j} \mathbf{A}^{(l)}_{kj}$ is the
\emph{out-strength} of node $k$ during time step $l$, so that
$\sum_{j} \mathbf{T}^{(l)}_{kj}=1$. A generic $M$-step path
$\mu$ between two nodes $I$ and $J$ is defined as a
$(M+1)$-uplet $\mu\equiv\bigl\{I, k_{1},\, ...\, , k_{M-1}, J
\bigl\}$ providing a sequence of nodes crossed to reach $J$ at
time $t_M$ from $I$ at time $t_0$. Under the Markovian
assumption the probability for a random walker to take the path
$\mu$ under the condition of starting at $I$ is:
\begin{equation}\label{mostprobpathprob}
(p^{M}_{IJ})_{\mu} =
\mathbf{T}^{(1)}_{Ik_{1}} \Bigg[
\prod_{l=2}^{M-1} \mathbf{T}^{(l)}_{k_{l-1}k_{l}}
\Bigg] \mathbf{T}^{(M)}_{k_{M-1}J} \ .
\end{equation}
The most probable path (MPP) $\eta_{IJ}^{M}$ is the path that
maximizes Eq. \eqref{mostprobpathprob} with respect to the
intermediate nodes $k_2,...,k_{M-1}$. Its probability is
denoted by $P^{M}_{IJ}=\max_{\mu}\{ (p^{M}_{IJ})_{\mu}\}$. The
exact maximization of Eq.~(\ref{mostprobpathprob}) can be obtained
iteratively by noting that in the first step the maximum
probability to reach a given node $k_1$ is simply
$P^1_{Ik_1}=\mathbf{T}^{(1)}_{Ik_1}$ and then using the
recurrence
\begin{equation}
P^{l+1}_{Ik_{l+1}}=\underset{k_l}{\mathrm{max}} \left( P^l_{I k_l} \mathbf{T}^{(l+1)}_{k_l k_{l+1}}\right)
\end{equation}
for $l=1,2,...,M-1$ until reaching $k_M=J$. This type of
iterative optimization is similar (taking logarithms of the
probabilities involved) to the one used to find optimal
configurations of directed polymers in random media
\cite{halpinhealy1995kinetic} and can be considered as an
adaptation of the classical Dijkstra algorithm
\cite{dijkstra1959note} to the layered and directed structure
of the TOG. The computational cost of the maximization is
strongly reduced by calculating first \textsl{accessibility
matrices} \cite{lentz2013unfolding} and restricting the
maximization search to the set of nodes that are {\sl
accessible} from $I$ and for which $J$ results {\sl accessible}
as well.
%

To assess whether the MPP alone is a good representation of the
transport dynamics we introduce the quantity  $\lambda_{IJ}^M
\equiv P^{M}_{IJ}/\sum_{\mu}(p^{M}_{IJ})_{\mu}$. It corresponds
to the fraction of probability carried by the MPP between $I$
and $J$ with respect to the sum of the probabilities of all
paths connecting these nodes after $M$ steps.
%
Note that the sum in
the denominator can be efficiently computed as the entry
$(I,J)$ in the matrix product
$\prod_{l=1}^{M}\mathbf{T}^{(l)}$. Depending on the network
under investigation, the MPP can actually carry a significative
fraction of the total connection probability. When this is not
the case we can relax the definition of MPP and define a subset
of HPPs which carry most of the probability. In particular we
want to identify paths characterized by individually carrying a
probability larger than a fraction $\epsilon$ of the MPP
probability. i.e. larger than $\epsilon P_{IJ}^M$, with $0\leq
\epsilon \leq 1$. Exhaustively searching for all such paths
becomes computationally prohibitive except for the smallest $N$
and $M$ values. Here we compute the set $\mathcal{Q}_{IJ}^M$ of
all paths of $M$ steps between $I$ and $J$ that are constructed
by joining the MPP from I to an intermediate $k_l$ and the MPP
from $k_l$ to $J$. In principle there would be $(M-2)\times N$
such paths, one for every choice of the intermediate $k_l$, but
this number is in fact much smaller when considering that $k_l$
should be accessible from $I$ \cite{lentz2013unfolding}, that $J$ should be accessible from $k_l$ itself, and that many of the resulting paths
turn out to be repeated. Out of these we consider the subset
$\mathcal{K}_{IJ}^M(\epsilon)=\left\{\mu \in \mathcal{Q}_{IJ}^M
~|~ (p^{M}_{IJ})_{\mu}
> \epsilon P_{IJ}^M\right\}$. This set contains the MPP, and although it may miss some of the
paths with probability larger than $\epsilon P_{IJ}^M$ we
expect it to contain a sufficiently representative sample of
them. This can be checked by calculating
$\lambda_{IJ}^M(\epsilon)\equiv\sum_{\nu
\in\mathcal{K}_{IJ}^M(\epsilon)}(p^{M}_{IJ})_\nu/\sum_{\mu}(p^{M}_{IJ})_{\mu}$,
the fraction of probability carried by this set of HPPs.


\section{MPP-betweenness}

Equipped with the above definitions we
can now characterize network properties that are dependent on
optimal paths in different ways. One of these is the concept of
\textsl{betweenness centrality}, which is generally
defined as the proportion of shortest paths passing through a
node. We introduce here a definition based on the number of
\emph{most probable paths} crossing a node. Specifically we
define the betweenness of node $K$ after $M$ steps as
$\mathcal{B}_{K}^{M} = \sum_{IJ} g_{IJ;K}^M/N_{M}$, where the
sum is over all pairs of initial nodes $I$ and final accessible
nodes $J$, $N_M$ is the total number of connected pairs of
nodes at time step $M$ (computable from accessibility matrices
\cite{lentz2013unfolding}), and $g_{IJ;K}^M$ is the number of
times the node $K$ appears in the most probable path connecting
$I$ and $J$.
Fixing the time interval $M$ corresponds to considering paths
with the same temporal duration. In this way we ignore
connections that are occurring at shorter or longer
times~\cite{kim2012temporal} and that can be significantly more
probable. It is possible to overcome this limitation by
performing a multistep analysis: we can look at all MPP's with
$M$ in a given interval $[M_{min},M_{max}]$ and choose the MPP,
$\eta_{IJ}^{[M_{min},M_{max}]}$, with the highest probability.
The multistep analysis leads to an alternative definition of
betweenness, i.e a multistep MPP-betweenness
$\mathcal{B}_{K}^{[M_{min},M_{max}]}$ which is calculated
considering the multistep MPPs instead of the fixed-$M$ one.

\begin{figure*}
\centering
\includegraphics[width=\textwidth, clip=true]{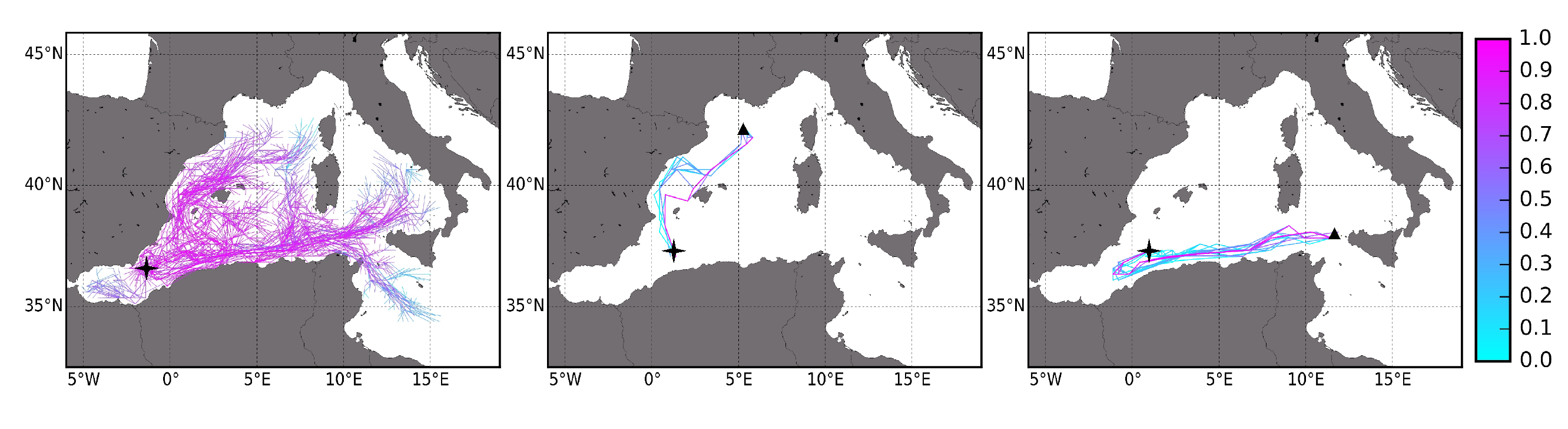}
\caption{(Color online) Paths of $M=9$ steps (three months) in the
Mediterranean flow network with starting date January 1st 2011,
represented as straight segments joining the path nodes. Left:
MPPs originating from a single node (black star) and ending in
all accessible nodes. Color gives the $P_{IJ}^M$ value of the
paths in a normalized log-scale between the minimum value
($10^{-15}$, light turquoise) and the maximum ($10^{-5}$, dark pink).
Center and right: all the paths in the
$\mathcal{K}_{IJ}^M(\epsilon)$ set with $\epsilon=0.1$, initial
point marked by a cross and final point marked by a triangle.
The center panel shows the 18 HPPs, out of a total of 54276
paths between the two sites. The MPP, with $P_{IJ}^M=3\times
10^{-9}$, is displayed in dark red, whereas the other paths are
colored with a normalized logarithmic scale according to their
$(p_{IJ}^M)_\mu$ values in $[\epsilon P_{IJ}^M,P_{IJ}^M]$.
Right panel shows the 39 HPPs, out of a total of $61\times
10^6$, in a similar logarithmic scale normalized in $[\epsilon
P_{IJ}^M,P_{IJ}^M]$ with $P_{IJ}^M=1.4\times 10^{-6}$. }
\label{fig:pathvisual}
\end{figure*}

\section{Mediterranean flow network}

We apply the previous
concepts to a specific network describing surface water
transport in the Mediterranean Sea
\cite{rossi2014hydrodynamic,sergiacomi2015flow}. Fluid
transport is typically studied from a Lagrangian perspective,
by following trajectories of particles released in the flow.
Recent works have approached the problem from a discretized
point of view~\cite{froyland2007detection,froyland2012three,
speetjens2013footprints,rossi2014hydrodynamic,sergiacomi2015flow}.
Most of these studies are focused on the analysis and
identification of coherent structures like vortices, barriers,
or regions where trajectories of fluid parcels tend to be
confined. Less insight is available on the pathways followed by
fluid masses during transport and in the resulting connectivity
patterns. In this regard, our approach complements the standard
Lagrangian toolbox as will be illustrated with this oceanic
flow example.

Velocity data have been collected from the Mediterranean
Forecasting System Model, Physics Reanalysis
component~\cite{oddo2009nested}. We use the horizontal surface
velocity daily field of the whole Mediterranean basin at a
resolution of $\sfrac{1}{16}$ degrees during $10$ years of
simulation (2002-2011). For the first months of each year a
temporal flow network has been constructed by partitioning the
surface of the Mediterranean sea in $N=3270$ two-dimensional
square-boxes (approximate lateral size of 28 km), each of which
is identified with a network node. Link's weights are assigned
from the effective mass transport driven by ocean currents
between two boxes during a given time
interval~\cite{rossi2014hydrodynamic,sergiacomi2015flow}. To
build the adjacency matrices at each time $t_l$ ($l=1...M$) we
homogeneously initialize $n=500$ ideal fluid particles in each
node and integrate the surface velocity for a given time
$\tau$. The matrix element $\mathbf{A}^{(l)}_{IJ}$ is the
number of particles starting from node $I$ at time $t_{l-1}$
and arriving to node $J$ at $t_l$. The normalized matrices
$\mathbf{T}^{(l)}$ define a flow network.

We perform the calculations using a time step of $\tau=10$
days, and considering $M$ from $6$ to $9$ steps (i.e. in
between $2$ and $3$ months, a time interval during which the
horizontal-flow assumption remains a good approximation).
We build 10 temporal networks, each one having $t_0$ as
January 1st of each of the years available in the simulation
database ($2002-2011$).

\section{Results}


In Fig.~\ref{fig:pathvisual} (left panel) we show on map the
set of all the MPPs of $M=9$ time steps starting from a given
node in one of our temporal networks (the one corresponding to
2011), and we notice how many different connections are
possible from a single starting node. The $P_{IJ}^M$ values
span several orders of magnitude and this behavior is typical
for the distribution of probability across MPPs. We stress here
that MPPs do not coincide in general with fastest paths: the
fastest connection among two nodes is not always the most
probable one (see Appendix) stressing the
importance of a weighted description of the network.

\begin{figure}
\centering
\includegraphics[width=1.0\columnwidth, clip=true]{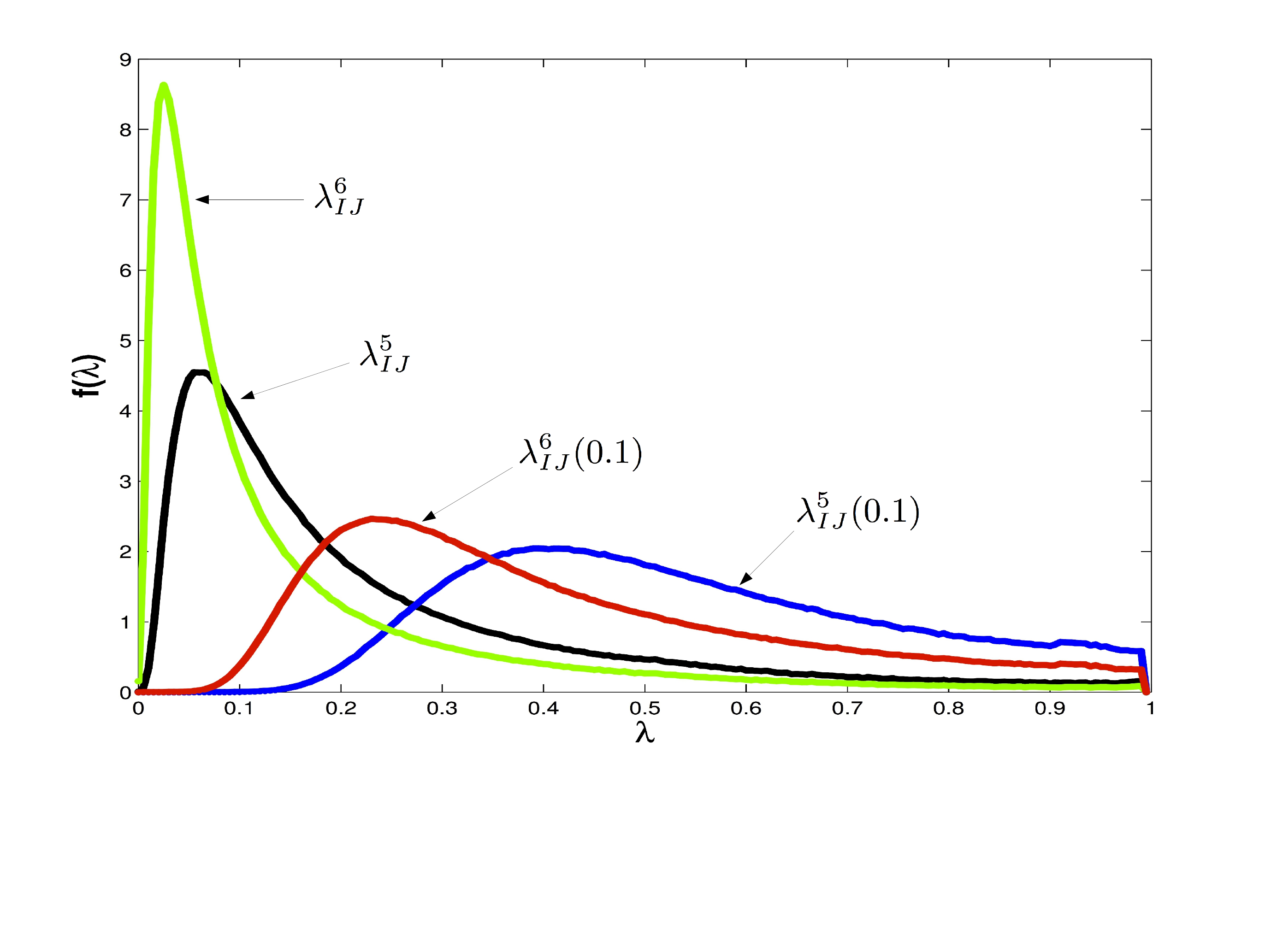}
\vspace{-50pt}
\caption{(Color online) Normalized histogram $f(\lambda)$ of coefficients $\lambda_{IJ}^M$ for $M=5$
(black curve) and $M=6$ (green curve), and
$\lambda_{IJ}^M(\epsilon)$, with $\epsilon=0.1$ for $M=5$ (blue curve) and $M=6$ (red
curve). The statistics is compiled from all connected pairs of nodes $(I,J)$ and the ten
temporal flow networks corresponding to the first months of the ten years of velocity data.
The mean values are: $\langle {\lambda_{IJ}^{5}} \rangle=0.24$;
$\langle {\lambda_{IJ}^{6}} \rangle=0.16$; $\langle {\lambda_{IJ}^{5}} (0.1) \rangle=0.52$;
$\langle {\lambda_{IJ}^{6}} (0.1) \rangle=0.42$. }
\label{fig:subdistr}
\end{figure}

To assess how representative of the whole dynamics are MPPs
such as the ones shown in Fig.~\ref{fig:pathvisual} we show in
Fig.~\ref{fig:subdistr} the distribution of $\lambda_{IJ}^M$
and $\lambda_{IJ}^M(\epsilon)$ for two values of $M$. The
distributions are collected from the $\lambda$-values of the
whole set of accessible pairs $(I,J)$ in our ten temporal
networks. For small $M$ most of the MPP have significant
$\lambda$-values, but as $M$ increases the peak in the
distribution of $\lambda_{IJ}^M$ shifts towards zero (we have checked that exponentially) as a
consequence of the dramatic increase with $M$ of the number of
available paths between two nodes. Then, it becomes important
to consider larger sets of paths such as
$\mathcal{K}_{IJ}^M(\epsilon)$. For the cases plotted, i.e.
$M=5,6$ and $\epsilon=0.1$, the mean values of
$\lambda_{IJ}^M(\epsilon)$ are around 0.5. This means that,
despite $\mathcal{K}_{IJ}^M(\epsilon)$ may not contain the full
set of paths which individually carry a probability larger than
$\epsilon P^M_{IJ}$, it is large enough so that, for most of
the $(I,J)$ pairs, it contains globally over $50\%$ of the
connection probability. This result further gains meaning if we
consider that the number of paths in
$\mathcal{K}_{IJ}^M(\epsilon)$ is on average well below $1\%$
of the total number of paths of $M=5$ and 6 steps. Hence,
despite the strong particle dispersion characterizing our flow
networks it is true that only a small subset of paths
contribute significantly to the main transport features. This
conclusion will also show up when studying global network
properties, such as the betweenness centrality measure.

\begin{figure}[h]
\centering
\includegraphics[width=0.5\textwidth, clip=true]{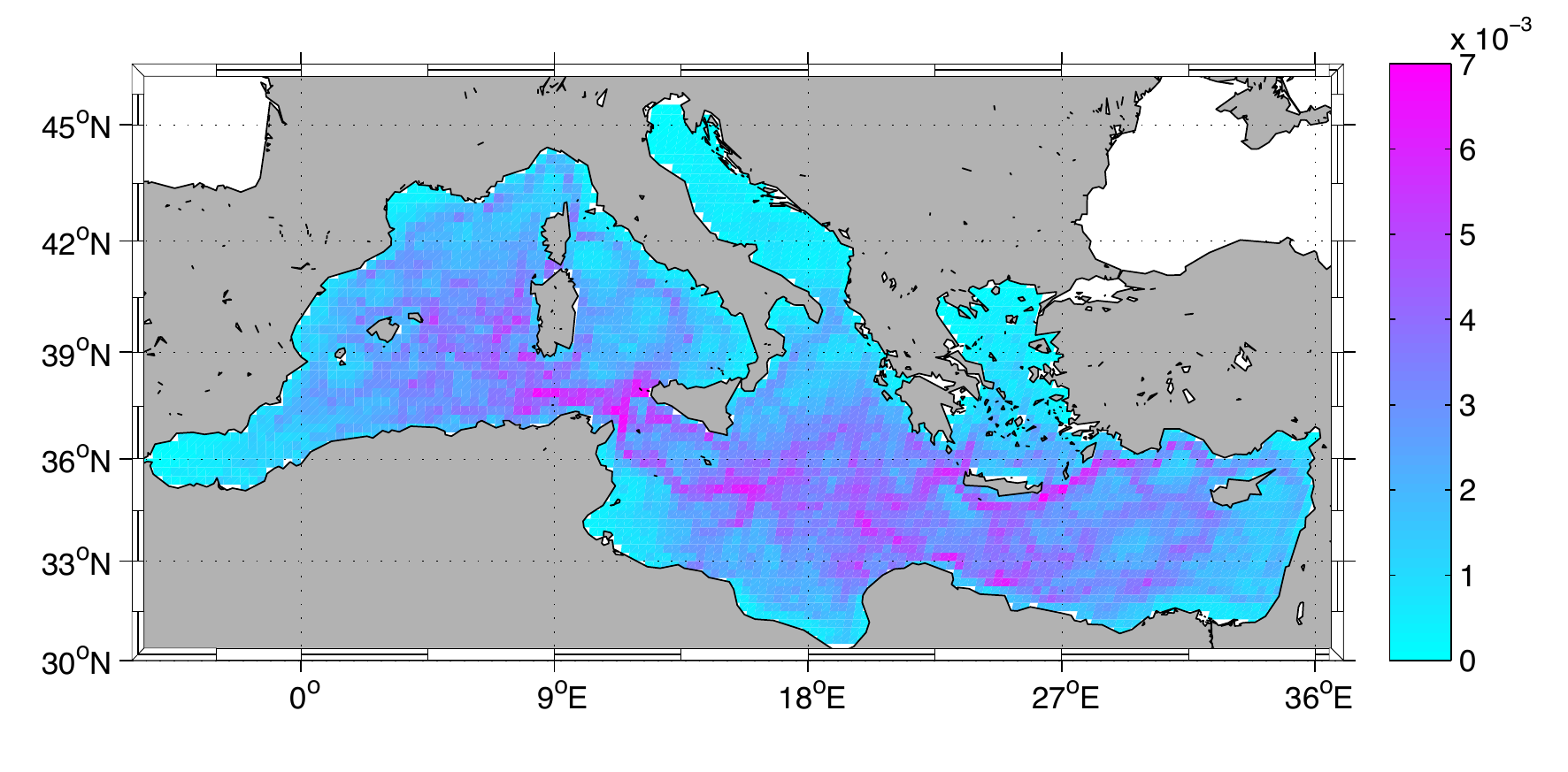}
\caption{(Color online) Multistep MPP-betweenness $\mathcal{B}_{K}^{[6,9]}$ at each geographical node $K$,
computed for each of our ten (2002-2011) temporal networks and then averaged over them.}
\label{fig:betweenness}
\end{figure}

In Fig.~\ref{fig:betweenness} we show the multistep
MPP-betweenness $\mathcal{B}_{K}^{[6,9]}$, averaged over our
ten networks. We have noticed that the distribution of the
betweenness decreases exponentially at large betweenness values
so that there are not strong hubs in the network. Spatial
patterns determined by the transport dynamics of the flow are
clearly evident in the figure where high betweenness areas are
organized in one dimensional-like structures corresponding to
the main corridors of transport, i.e. narrow pathways that
connect different regions of the ocean. Main paths of the
Mediterranean sea are observed like Cyprus and Rhodes Gyres,
surrounding the Ionian basin, the Algerian current and those
along the Sicily strait, etc. Note that because of the
ten-years average, individual short lived mesoscale features
(eddies and fronts) are averaged out. We note however that,
despite this inter-year average, MPP-betweenness maps remain
dependent on the starting date $t_{0}$. These observations
highlight the seasonal and inter-annual variability of the flow
and justifies further the need of a time-dependent network
approach as opposed to a fully averaged static network
description.

The robustness of our methodology has been tested by checking
the stability of our results on MPP-betweenness under different
conditions. First, dealing with temporal networks it is
important to understand how much the results are affected by the
choice of the the time-step duration $\tau$
~\cite{ribeiro2013quantifying}. We checked this issue
considering different $M$ and $\tau$ values but the same total
duration. Results for the MPP-betweenness remain nearly
unchanged when keeping a constant total duration $M\tau$,
confirming robustness under variations of the temporal
resolution. Second, we noted that the MPP-betweenness does not
significantly change when computed with just the $50\%$ of MPPs
with larger values of $\lambda$ (i.e. when using a threshold to
retain only the most significant MPPs). Finally, to support our
interpretation of most probable paths as main carriers of
connectivity, we considered also betweenness calculated from
HPP subsets of, i.e. $g_{IJ;K}^M$ is now the number of times
node $K$ appears in the set $\mathcal{K}_{IJ}^M(\epsilon)$ of
HPPs between $I$ and $J$, with $\epsilon=0.1$. We did not
appreciated relevant differences between this calculation and
the one involving only MPPs. Indeed the Pearson correlation
coefficient between the two betweenness is larger than $0.9$.
Hence, despite the  MPPs represent a small portion of the paths
in the $\mathcal{K}_{IJ}^M(\epsilon)$ subsets (between $3-10\%$
for $\epsilon=0.1$, depending on the value of $M$), which is
itself a very small fraction of the full set of paths in the
network, they seem to be representative of the main
spatio-temporal structures describing the global dynamics.
Indeed, center and right panels of Fig. \ref{fig:pathvisual}
show that most of the relevant paths remain spatially close to
the MPP. This observation is confirmed by calculations of the
spatial dispersion between paths in
$\mathcal{K}_{IJ}^M(\epsilon)$, whose average turns out to be
of the order of the size of the boxes defining the nodes.

\section{Conclusions}

In this paper we introduced tools to compute
highly probable paths in weighted temporal networks and to
evaluate their relative importance. Betweenness centrality
measures based on them have also been introduced.
We applied this approach to characterize connectivity in the
Mediterranean Sea from a network-theory perspective.
Here, MPPs correspond also to the set of paths that maximize
the fraction of transported mass, giving therefore a clear
physical interpretation of connection probabilities.
Despite MPPs represent only a small fraction of the whole set
of paths, we found that they suffice to highlight the main
transport pathways across our network, since most of
the HPPs remain geographically close to them. This means that paths followed by fluid masses connecting two regions are organized in elongated narrow tubes centered on the MPP.

We believe that the study of fluid transport as a network will
provide new tools and insights complementing standard
Lagrangian methods.
Indeed most of these are devoted to the identification of
barriers to transport or coherent regions with small fluid
exchange with the surroundings. Here we are instead addressing
the opposite question: how to detect regions and pathways that
maximize fluid interchange across the network. Even
though in principle pathways are simply given by trajectories,
it is almost impossible to extract clear and significant patterns
from the complex sets of trajectories that arise in all, except
the most simple, time-dependent flows. Our approach allows to quantify explicitly transport
among two sub-regions of the domain, highlighting the optimal path connecting them.
In this sense MPP-analysis differs from simply studying
the evolution in time of tracer concentrations seeded in
a given region.

Beyond the fluid dynamics context, MPPs and the MPP-betweenness
measure here introduced could be easily transferred to other
kinds of weighted temporal networks. This could be relevant,
for instance, in defining vulnerability metrics in disease
spreading processes, individuate critical nodes in
biological/ecological networks or in detecting bottlenecks of
reaction chains in metabolic networks.

\acknowledgments
We acknowledge financial support from FEDER
and MINECO (Spain) through the ESCOLA (CTM2012- 39025-C02-01)
and INTENSE@COSYP (FIS2012-30634) projects, and from European
Commission Marie-Curie ITN program (FP7-320 PEOPLE-2011-ITN)
through the LINC project (no. 289447). We acknowledge helpful
discussions with P. Fleurquin and V. Rossi.

\appendix*
\section{Comparing fastest and most probable paths}

In the study of temporal networks, the concept of
\emph{fastest} path has been put forward as a natural extension
of the \emph{shortest path} of static networks. In our work we
define and analyze a different type of relevant path which is
the Most Probable Path (MPP). It is important to address the
differences between most probable and fastest paths, and we do
so in this Appendix.

The MPP refers to the path transporting the maximum fraction of
water (or of probability) between two nodes, and the fastest
path to the pathway linking the two nodes in the shortest time.
This second concept can not be implemented when the number of
time-steps is fixed. However we can reclaim the concept of
fastest path within a multistep approach, i.e. by looking at a
time window specified by a range of values for the number of
time steps $M$. We can then define the set
$\mathcal{M}_{IJ}^{[M_{min},M_{max}]}$  of $(M_{max} - M_{min}
+ 1)$-MPPs for the pair $I,J$ for  $M\in [M_{min},M_{max}]$,
and the \emph{fastest}-MPP as the MPP in
$\mathcal{M}_{IJ}^{[M_{min},M_{max}]}$ corresponding to the
smaller $M$. On the other side we can also define an
\emph{absolute}-MPP, i.e. the MPP in
$\mathcal{M}_{IJ}^{[M_{min},M_{max}]}$ having the highest
probability. By comparing the set of \emph{absolute}-MPPs with
the set of \emph{fastest}-MPPs we can address the question: is
the fastest path necessarily the most probable?
\begin{figure}[h]
\centering
\includegraphics[width=0.5\textwidth, clip=true]{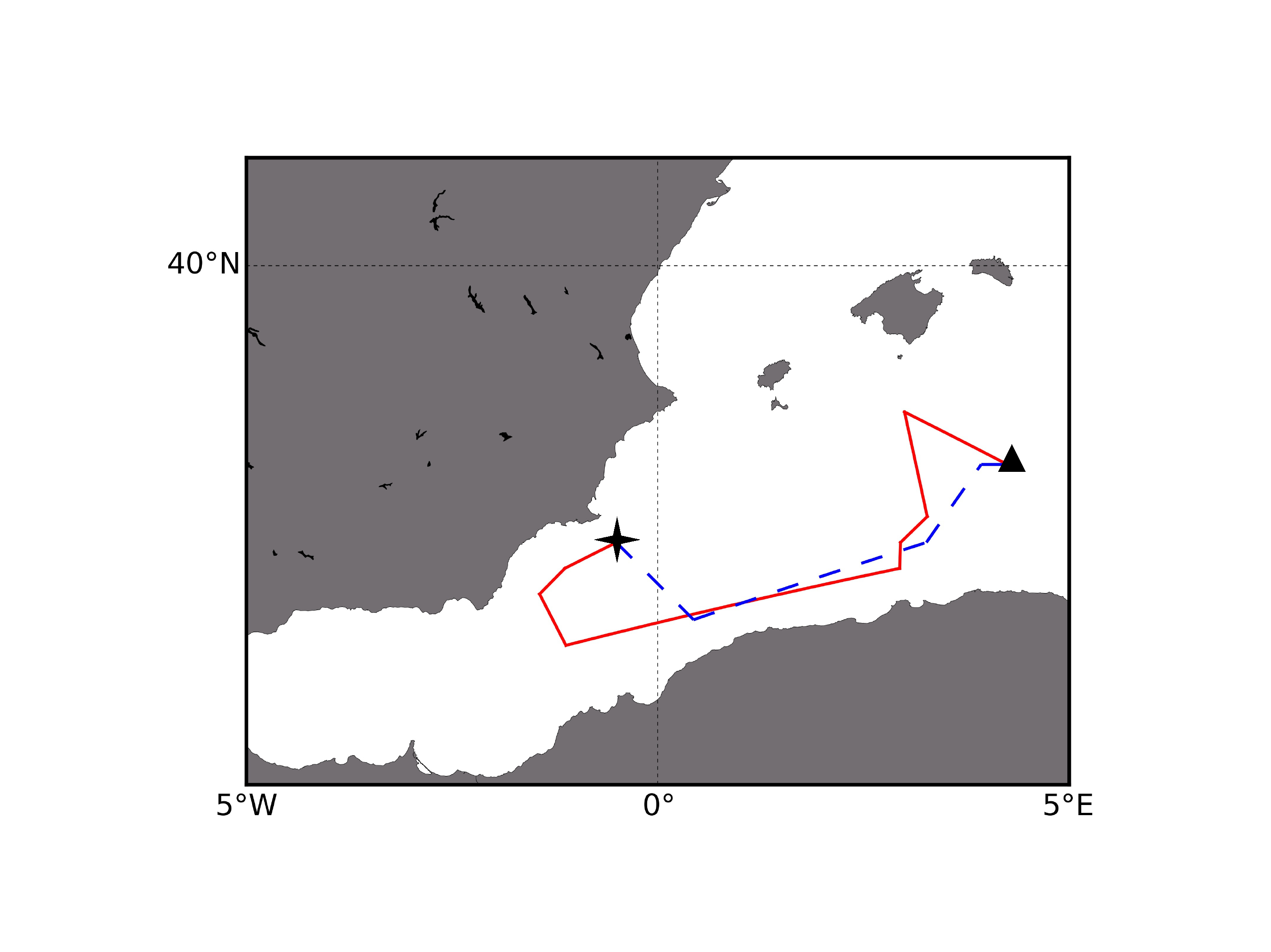}
\caption{(Color online) We show the \emph{fastest}-MPP (dashed blue line) and the \emph{absolute}-MPP (continuous red line), between an origin node $I$ (black star) and a destination node $J$ (black triangle). The considered full set of paths ranges The \emph{fastest}-MPP reaches the destination node in $4$ steps of $\tau=10$ days while the \emph{absolute}-MPP needs $8$ steps i.e. $40$ days more. The probability associated to the \emph{fastest}-MPP is $5.9 \times 10^{-7}$ and the probability  of the \emph{absolute}-MPP is $6.7 \times 10^{-6}$.}
\label{paths_fastestvsmpp}
\end{figure}

In Fig.~\ref{paths_fastestvsmpp} we show that the
\emph{fastest}-MPP  among two nodes of the network is different
to the \emph{absolute}-MPP and that its probability, in several
cases, can be orders of magnitude smaller. We considered for
this example paths ranging from $M = 3$ to $M=9$ steps of $10$
days (i.e. a time scale of $1-3$ months) with starting date
January 1st 2011. The results show the importance to
distinguish between the connections realized in the shortest
time and the connections that carries most of the transported
mass (the most probable).

To display in a more systematic way the differences between
\emph{fastest} and \emph{absolute} MPPs across the network we
study the rank plot of the whole set of paths during ten years
($2002-2011$) ranging from $3$ to $9$ steps of $10$ days
starting at January 1st of each year. The rank plot displays
the probabilities of each path in one of the sets sorted in
decreasing order. We see a gap in probabilities between the two
sets of about one order of magnitude in most of the range
displayed. The \emph{fastest}-MPPs have probabilities
significantly smaller than \emph{absolute}-MPPs.
\begin{figure}
\centering
\includegraphics[width=0.5\textwidth, clip=true]{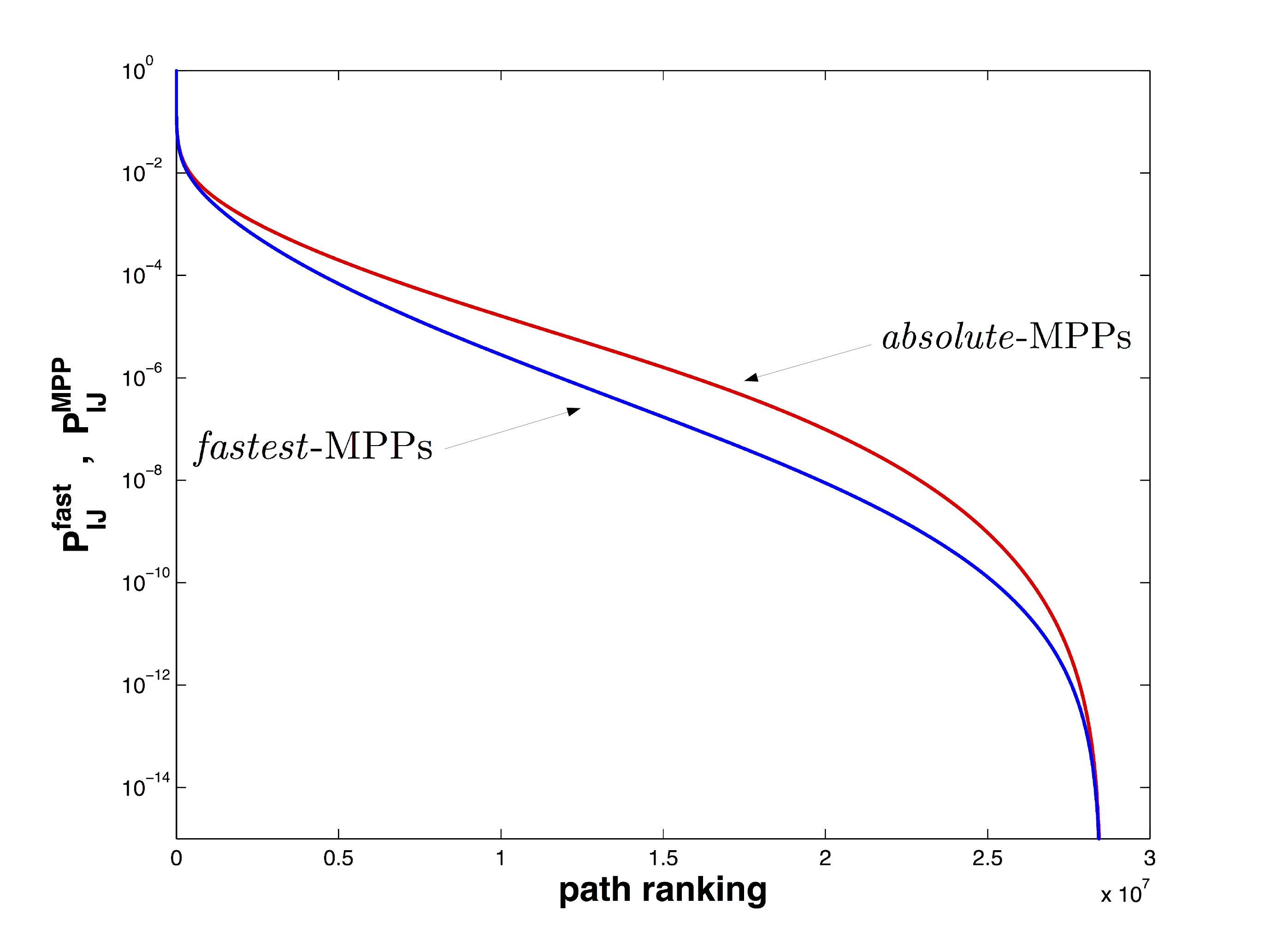}
\caption{(Color online) Ranking plot in which $P_{IJ}^{fast}$ (blue line) correspond to the probability of \emph{fastest}-MPPs and $P_{IJ}^{MPP}$ (red line) correspond to the probability of \emph{absolute}-MPPs. The range of probability values can be read from the vertical axis (logarithmic scale).  The total number of optimal paths can also be read-off from the horizontal axis.}
\label{rank_fastest_mpp}
\end{figure}

Finally, we also evaluate how these differences are reflected
in the betweenness measures. We define the relative difference
among the betweenness computed using the set of
\emph{fastest}-MPPs and \emph{absolute}-MPPs for the node $K$
as:
\begin{equation}
\Delta_{K} = 2 \frac{\mathcal{B}_{K}^{abs} -  \mathcal{B}_{K}^{fast} }{ \mathcal{B}_{K}^{abs} + \mathcal{B}_{K}^{fast} },
\end{equation}
where $\mathcal{B}_{K}^{abs}$ is the betweenness computed using
\emph{absolute}-MPPs and $\mathcal{B}_{K}^{fast}$ the
betweenness computed using  \emph{fastest}-MPPs. We consider
once more paths ranging from $3$ to $9$ steps of $10$ days with
starting date January 1st 2011 and we compute the
spatial-average (i.e. the average over nodes $K$) for the
absolute value of the relative difference finding $\langle
|\Delta_{K}| \rangle_{K} = 0.32$. This means that, on average,
the difference between the two measures is about $30\%$.

We stress that all the comparisons above are among paths that
are already MPPs linking a pair of nodes. Considering still
fastest paths (for example the one by which the very first
particle from one node reaches the other) will lead to much
stronger differences. In summary, the results show the
importance to distinguish between the connection realized in
the shortest time and the connection that carries most of the
transported mass. This gains even more relevance when
considering possible applications such as, rescue operations,
pollutant-spreading or biological connectivity.

\end{document}